\documentclass[aps,pra,showpacs,amsmath,twocolumn,superscriptaddress]{revtex4-1}
\usepackage{amsmath,amssymb,graphicx}
\usepackage[final]{hyperref}
\usepackage{booktabs}

\hypersetup{
	colorlinks=true,       % false: boxed links; true: colored links
	linkcolor=blue,          % color of internal links
	citecolor=blue,        % color of links to bibliography
	filecolor=magenta,      % color of file links
	urlcolor=blue
	}

\begin{document}

\title{Multi-pass configuration for Improved Squeezed Vacuum Generation \\ in Hot Rb Vapor}

\author{Mi Zhang}
\author{Melissa A. Guidry}
\affiliation{Department of Physics, College of William $\&$ Mary, Williamsburg, Virginia 23187, USA}
\author{R. Nicholas Lanning}
\author{Zhihao Xiao}
\author{Jonathan P. Dowling}
\affiliation{Hearne Institute for Theoretical Physics and Department of Physics $\&$ Astronomy, Louisiana State University, Baton Rouge, Louisiana 70803, USA}
\author{Irina Novikova}
\author{Eugeniy E. Mikhailov}
\email[eemikh@wm.edu]{}
\affiliation{Department of Physics, College of William $\&$ Mary, Williamsburg, Virginia 23187, USA}

\date{\today}

\begin{abstract}
We study a squeezed vacuum field generated in hot Rb vapor via the polarization self-rotation effect. 
Our previous experiments showed that the amount of observed squeezing may be limited by the contamination of the squeezed vacuum output with higher-order spatial modes, also generated inside the cell. 
Here, we demonstrate that the squeezing can be improved by making the light interact several times with a less dense atomic ensemble. With optimization of some parameters we can achieve up to $-2.6$~dB of squeezing in the multi-pass case, which is $0.6$~dB improvement compared to the single-pass experimental configuration. 
Our results show that other than the optical depth of the medium, the spatial mode structure and cell configuration also affect the squeezing level.

\end{abstract}

%\keywords{}

\maketitle

\section{Introduction}

An increasing number of applications, such as quantum information~\cite{DLCZ,lukin03rmp,kimbleNature08,polzikRMP10,QuantMemoryJMO2016} and quantum sensor technologies~\cite{bachor_guide_2004,budker_optmagn_book,vanier05apb,mitchel2010prl_sqz,polzik2010prl,
mikhailov2012sq_magnetometer,otterstrom2014SelfSqMag}, rely on strong coupling of optical fields to long-lived atomic spin states. Here, we concentrate on the nonlinear process of polarization self-rotation (PSR)~\cite{Davis:92,budkerPRA01,novikova02JMO} in atomic vapor, which gives rise to a squeezed-vacuum generation, known as PSR squeezing~\cite{matsko_vacuum_2002,ries_experimental_2003, mikhailov2008ol,lezama2011pra,grangier2010oe,mikhailov2012sq_magnetometer}. In thermal atomic ensembles, it is expected that the strength of the interaction, and thus the overall performance increases at higher optical depth~\cite{lukin97prl,wynands'99,lukin03rmp,matsko_vacuum_2002,mikhailov2009jmo}. In practice, however, once an optimal value of an optical depth is reached, further increase leads to deterioration of the desired outcome, often together with increasing optical noise. Such behavior of PSR squeezing has been observed by several experiments~\cite{ries_experimental_2003,mikhailov2013ol_vortex}, and yet failed to be reproduced by the simple PSR squeezing theory, even if it included some known negative density-dependent effects (e.g., spin-exchange collisions, radiation trapping, etc)~\cite{matsko_vacuum_2002,mikhailov2009jmo}. Recently, we have demonstrated that worse-than-expected experimentally observed PRS squeezing can be caused by the modification of spatial mode decomposition of both squeezed vacuum and pump fields as a results of interaction with the atoms~\cite{MiZhangPRA16}. Similar multimode generation has also been reported in the non-degenerate four-wave mixing processes~\cite{LettPRL08,lettSci08,polzik2009oe,MarinoEJPD2012,marinoPRA16}.

In this paper we investigate the optical depth dependence of PSR squeezing using two methods for changing the optical depth: by changing the atomic density using the temperature of the cell, and by varying the length of the interacting atomic medium. Specifically, we compared the maximum amount of squeezing achievable when the laser beam traveled through a vapor cell once, or several times at different cell temperatures. We also let the light pass through two independent identical Rb vapor cells, optimizing the experimental conditions in each independently. In both cases we observed higher amount of squeezing for the case of light traversing the cell twice at lower temperature, even though the overall value of optical depth in both cases were comparable. A possible explanation for our results is consistent with the theoretically predicted non-trivial atomic density scaling of higher-order spatial mode generation, namely that less higher-order spatial modes are excited when the pump beam travels through a longer cell length at lower temperature than when it interacts with shorter atomic ensemble at higher atomic density.

%\section{Setup}

\begin{figure}[b]
	\includegraphics[width=1.0\columnwidth]{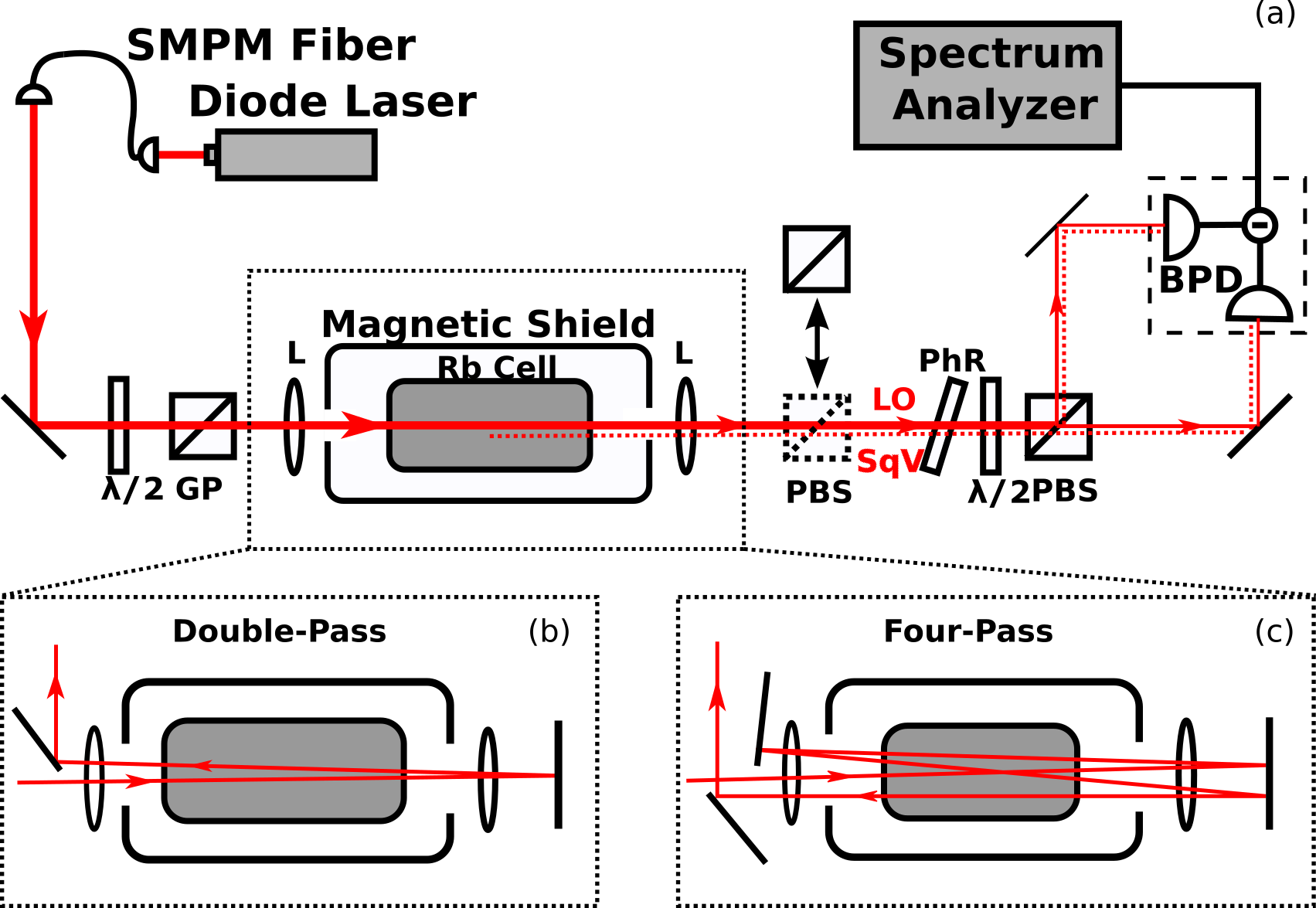}
		\caption{
		\label{fig:setup}(Color online)
		(a) Experimental setup for generation and measurement of PSR quantum noise in a single-pass configuration.
		SMPM is a single-mode polarization-maintaining fiber,
		$\lambda/2$ is half-wave plate,	
		GP is Glan-laser polarizer,
		PBS is a polarizing beam splitter,
		PhR is a phase-retarding wave plate,
		and BPD is a balanced photodetector.
		Insets (b) and (c) show the modification of the experiment for the double- and quadruple-pass arrangements, correspondingly.
	}
\end{figure}
 \begin{figure*}[t!]
	\includegraphics[width=1.0\textwidth]{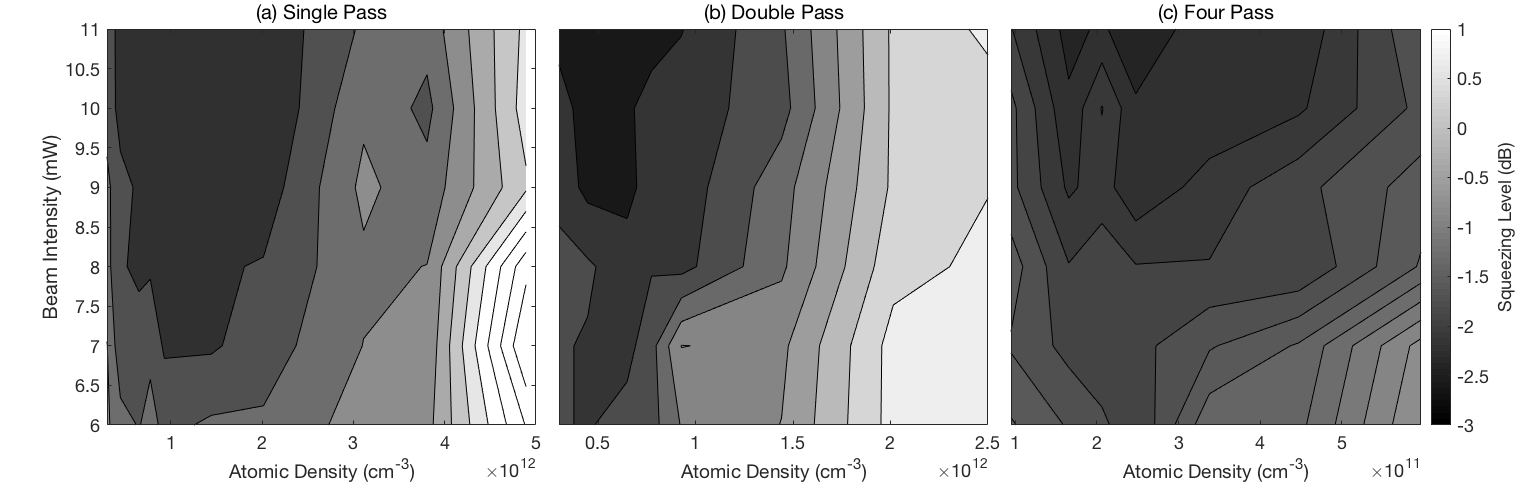}
	\caption{
		\label{fig:multi_pass_map}
		Experimental optimization of squeezed vacuum in the parameter space  of the pump beam power and atomic density for \textit{(a)} single, \textit{(b)}double and \textit{(c)} quadruple-pass configurations. The color represents the value of the minimum quantum noise suppression below the shot noise level (see the scale on the right), with the darkest areas in each plot representing the ``island'' of optimal conditions for squeezing.  Note the difference in horizontal scaling between the three graphs.
		}
\end{figure*}
%
%\begin{figure}[h]
%        \includegraphics[width=1.0\columnwidth]{ready_plots/double}
%        \caption{
%        \label{fig:double_pass}        
%        double pass}
%\end{figure}
%
%\begin{figure}[h]
%        \includegraphics[width=1.0\columnwidth]{ready_plots/four}
%        \caption{
%         \label{fig:four_pass}        
%        four pass}
%\end{figure}

\section{Multi-pass experimental arrangement}

The simplicity of the experimental arrangement is one of the attractive features of PSR squeezing generation. Before the cell, the optical input consists of a strong coherent field in one linear polarization, and a coherent vacuum in the orthogonal polarization. The strong field acts like a pump, preparing atoms in the necessary Zeeman superposition and thus modifying the quantum state in the orthogonally-polarized vacuum field, that becomes quadrature squeezed at the output of the cell.
The experimental setup for squeezing generation is shown in Fig.~\ref{fig:setup} and is similar to one previously used ~\cite{mikhailov2013ol_vortex,MiZhangPRA16}. We used an external cavity diode laser tuned at the $5^{2}S_{1/2}, F=2 \rightarrow 5^{2}P_{1/2},F^{\prime} = 2$ transition of ${}^{87}$Rb ($\lambda \simeq 795$~nm). 
Since in this experiment we paid special attention to the transverse mode structure of the optical fields, the output beam was spatially filtered using a single-mode-polarization-maintaining (SMPM) fiber to ensure its Gaussian transverse intensity profile. After passing a Glan laser polarizer to ensure the purity of its linear polarization, the beam was focused into a $7.5$~cm long cylindrical Pyrex cell containing isotopically enriched $^{87}$Rb vapor without any buffer gas. The waist of the focused beam (diameter $100~\mu$m at $1/e^2$ intensity level) was located $6.5$~cm from the front of the cell. The cell was mounted inside three layers of $\mu$-metal magnetic shielding, and its temperature was actively stabilized between room temperature and $120^{\circ}$C, that corresponds to a variation of Rb density between $1.1\times10^{10}$~cm$^{-3}$ and $2.0\times10^{13}$~cm$^{-3}$.

To establish the double-pass setup, we placed a mirror after the output collimating lense  to back-reflect the beam into the cell, as is shown in Fig.~\ref{fig:setup}(b). The four-pass case, shown in Fig.~\ref{fig:setup}(c), is achieved by adding an extra retro-reflecting mirror before the first lens, once more doubling the optical pass of the beam through the cell. It is important to note that for the double-pass configuration it was possible to fine-tune the focussing of the retro-reflected beam inside the cell by adjusting the position of the output lens, no additional beam control was available for the four-pass arrangement, and as a result the final output beam expanded compare to the single- or double-pass outputs.

To measure the quadrature noise in the squeezed field, we mix the the squeezed vacuum field and the pump field (that serves as a local oscillator) at a polarized beam splitter (PBS), oriented by $45^\circ$ with respect to the polarization directions of the two optical fields, after changing the phase difference between them with a phase-retarding (PhR) plate (a birefringent quarter-wave plate with a crystal axis set parallel to the LO polarization). The differential output of the balanced photodetector is analyzed using a spectrum analyzer, for which the detection frequency is fixed at $800$~kHz with a resolution bandwidth of $30$~kHz, and video resolution  of $100$~Hz. 

When necessary, one can reject any atom-induced quantum state modifications in the squeezing channel, and replace it with a coherent vacuum by inserting in the beam path a polarizing beam splitter (PBS) that transmits only the pump field.  We use this method, for example, to calibrate the shot noise level in our experiment. 

%The noise suppression in such setup is $2.1\pm0.2$~dB below the SQL level in the most squeezed quadrature and $11$~dB above shot noise in the orthogonal quadrature.

%\section{Single pass vs. multiple pass configuration comparison}
%

% single pass, two pass and four pass
We carried out a comparison of squeezing level in the three case, namely directing the pump beam through the vapor cell once (single pass), two times (double pass) or four times (quadruple pass). Since previous studies found that the best squeezing required careful optimization of both atomic density and pump laser power, we carefully mapped the measured minimum quantum noise as a function of both parameters for each experimental arrangement, as shown in Fig.~\ref{fig:multi_pass_map}.
Fortunately, the highest beam power ($11$~mW) yielded the best squeezing for each configuration, conveniently allowing us for the further measurements to fix the laser power at that value, and to confidently compare the squeezing measurements performed at different atomic densities. 
In case of a single pass, the best value of squeezing measured was $-2.0$~dB at the Rb density of $N^{\mathrm{opt}}_{\mathrm{sngl}} = 9.3 \times10^{11}$~cm$^{-3}$. 
When the optical path was doubled, a squeezing level of $-2.6$~dB was achieved at the atomic density $N^{\mathrm{opt}}_{\mathrm{dbl}} = 4.3\times10^{11}$~cm$^{-3}$, showing a substantial improvement over the single-pass case.  
For the quadruple-pass setup we saw a slight decrease in the measured squeezing amount to $-2.2$~dB at $2.4\times10^{11}$~cm$^{-3}$. However, as noted before, in this configuration we did not have full beam-shape control, and thus the expansion of the laser beam on the subsequent passes may have led to the lower observed squeezing.   

\begin{figure}[h!]
	\includegraphics[width=1.0\columnwidth]{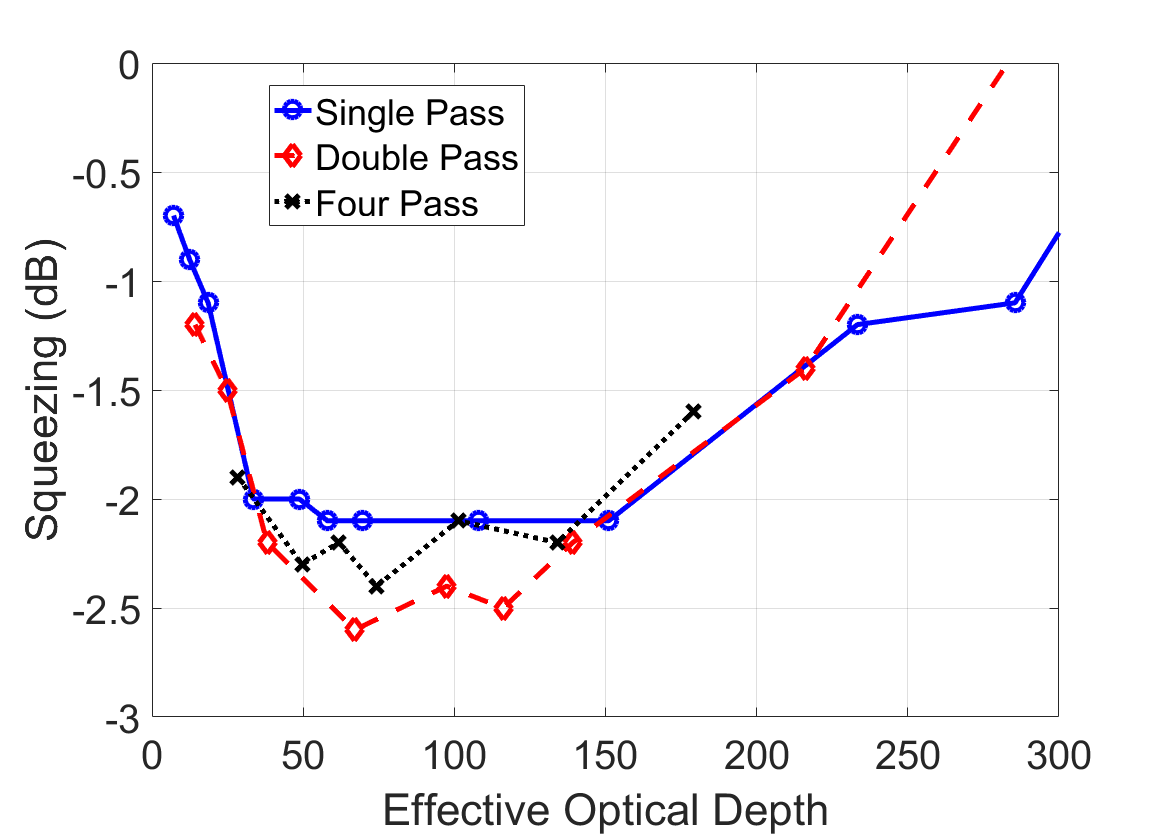}
	\caption{(Color online) Measured dependencies of minimum quadrature noise on the effective optical depth for single, double- and quadruple-pass configurations. The laser pump power for all three traces was $11$~mW,shown in Fig.~\ref{fig:multi_pass_map} to provide the best value of squeezing for all three geometries.
		\label{fig:effective}
}
\end{figure}

To highlight the different outcomes in case of the increased geometrical optical path and increase atomic density, Fig.~\ref{fig:effective} plots the values of measured squeezing as functions of an effective optical depth, defined as $d = \sigma_0 N L$, where $\sigma_0 = 10^{-9} \mathrm{cm}^2$ is the resonant absorption cross-section, $N$ is the atomic number density, and $L$ is the total length of the atomic medium (i.e. it is equal to the twice the length of the vapor cell for the double-pass arrangement). If the nonlinear interaction leading to vacuum squeezing generation depended only on the integrated optical depth of the ensemble, one could naively expect that all three curves would appear on top of each other. 
Indeed, there are several clear similarities between the three curves. The overall trends agree quite well: at the low-density limit the measured squeezing improved with optical depth, reaching a rather broad plateau; further increase in atomic density resulted in rapid deterioration of squeezing. In each configuration the best squeezing occurred at approximately the same optical depth range. The main difference between the three geometries appeared in the squeezing magnitudes reached in each case, with the single-pass configuration showing the worst quantum noise suppression ($-2.0$~dB), and the double-pass - the best ($-2.6$~dB).
This difference can be explained if we analyze the dependence of higher-order spatial mode generation during the light-atom interaction. As we demonstrated previously~\cite{MiZhangPRA16}, the presence of these higher-order modes can deteriorate the measured value of squeezing by two factors. First, the difference in the mode decomposition between the pump field (that serves as a local oscillator for our detection scheme) and the vacuum field reduces the effectiveness of homodyne detection. Second, the difference in quantum noise modifications  for different spatial modes result in mismatch between the detection quadrature providing best value of squeezing for different modes, and overall reduction of detectable squeezing when the mixture of spatial modes reaches the detector. 

\begin{figure}[h]
	\includegraphics[width=1.0\columnwidth]{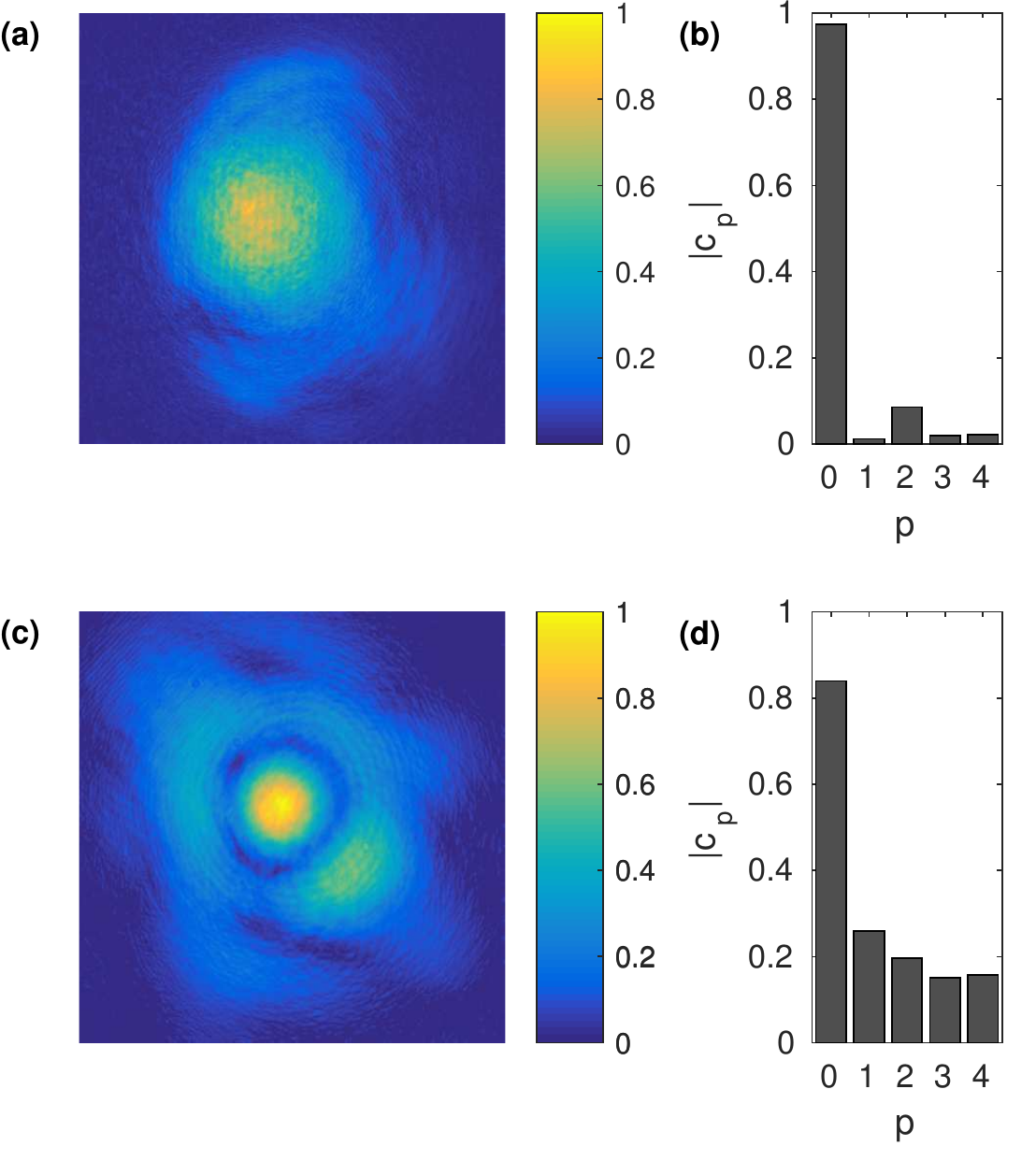}
	\caption{
		\label{fig:pumpmodesexpt}(Color online) Transverse intensity distribution of the pump field \textit{(a)} and the absolute values of the calculated deconvolution coefficients $|c_p|$ of the Laguerre-Gausse modes $L_p^{\ell=0}$, defined in Eq.(\ref{decomp}) after the cell at room temperature ($T=26^\circ$C, $N=10^{10}\mathrm{cm}^{-3}$). Small beam contamination due to diffraction give rise to non-zero $c_2$ coefficient, even if atoms are not present. Corresponding results for the pump beam after the interaction with Rb vapor at high temperature ($T=91^\circ$C, $N=2.6\times 10^{12}\mathrm{cm}^{-3}$) are shown in \textit{(c)} and \textit{(d)}. }
\end{figure}

The example of higher-order mode generation in the pump field is shown in Fig.~\ref{fig:pumpmodesexpt}. As expected, for lower atomic density the intensity distribution of the pump beam is well approximated by the the fundamental Gaussian mode $p=0$. However, if we operate the experiment at the higher than optimal atomic density, we can start observing clear signs of a more complicated transverse structure in the output pump beam. Namely, Fig.~\ref{fig:pumpmodesexpt}(c) shows a ring-like formation, characteristic to the higher-order Laguerre-Gauss modes. This behavior is even clearer in the mode-decomposition diagrams, shown in Fig.~\ref{fig:pumpmodesexpt}(b,d), in which the measured transverse intensity distribution  of the pump field after the vapor cell $I_{\mathrm{out}}(r)$ is approximated by a linear combination of Laguerre-Gausse modes $L_p^{\ell=0}$:
\begin{equation} \label{decomp}
I_{\mathrm{out}}(r) = I_0 \sum_{p=0}^{4}c_p L_p^{\ell=0}(r/w_0),
\end{equation}
where $I_0$ is an overall normalization parameter, $c_p$ are the decomposition coefficients ($\sum_{p=0}^{4}|c_p|^2 = 1$), $L_p^{\ell}(x)$ is an associate Laguerre-Gauss function, and $w_p$ is the measured waist of the incoming laser beam. Note that no higher-order $\ell$ modes are expected due to the conservation of the optical angular momentum in SPR process.
It is clear that even though the fundamental Gaussian mode dominates, the contribution of the higher-order $p$ modes increases dramatically with the temperature.  

\begin{figure}[h]
	\includegraphics[width=0.8\columnwidth]{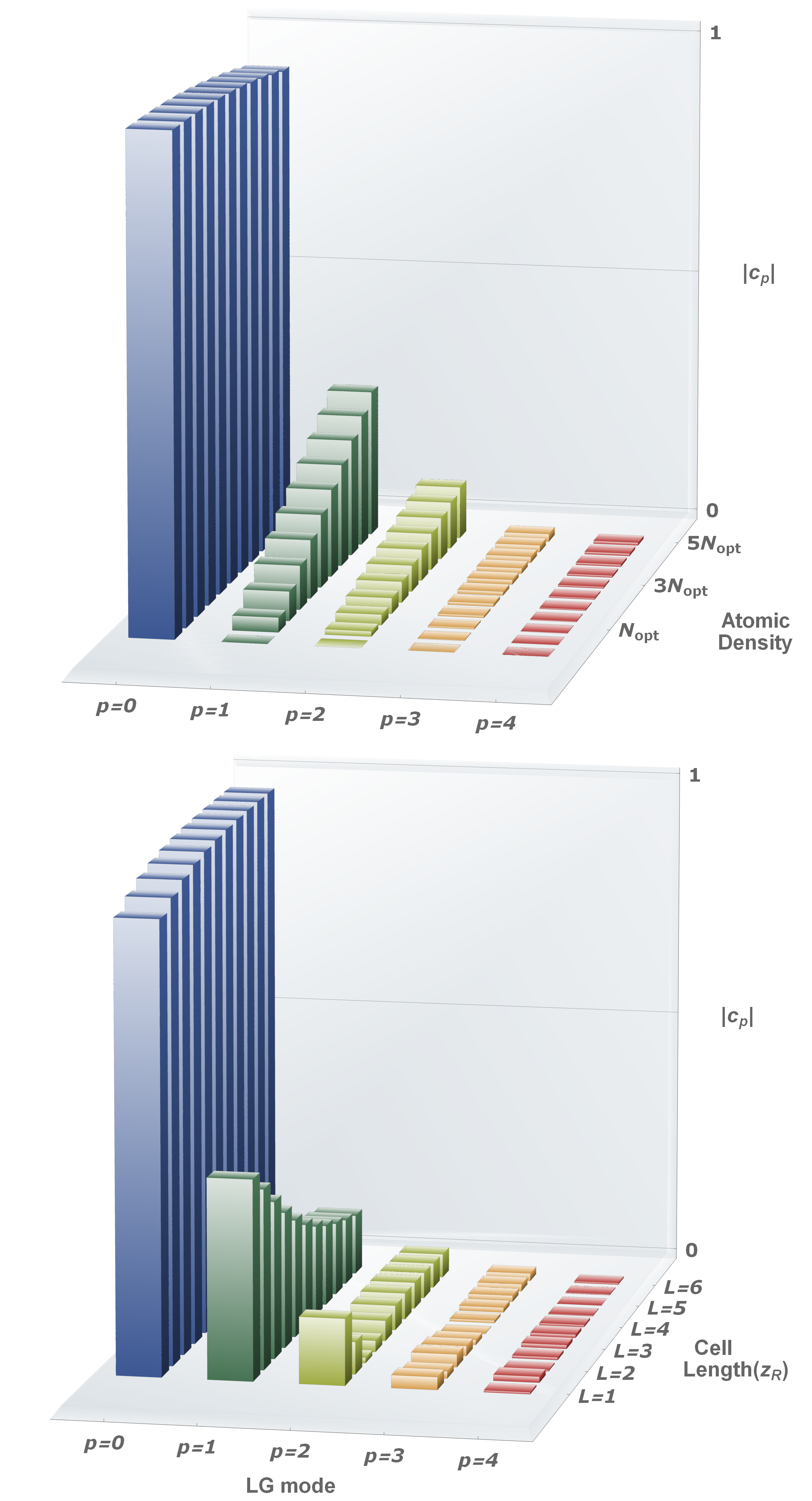}
	\caption{
		\label{fig:lengthvsdensitytheory}(Color online) The calculated relative amplitudes of the first few higher-order Laguerre-Gausse modes $L_p^{\ell=0}$ as \textit{(a)} atomic density $N$ gradually increases; or as \textit{(b)} the length of the atomic medium $L$ (in the units of the Rayleigh radius $z_R$) is increased.}
\end{figure}

Similar behavior is also predicted by the theoretical calculations using the method developed in Ref.\cite{nicktheorypaper}. Fig.~\ref{fig:lengthvsdensitytheory} tracks the change in the amplitudes of the generated higher-order modes $|c_p|$ for $p=0$ to $p=4$ in the cases the number density of atoms increases or the length of the atomic medium increases. 
In Fig.~\ref{fig:lengthvsdensitytheory}(a) we plot the output mode structure of the pump field at the output of the vapor cell for several values of atomic density $N$. The first step in the plot corresponds to no atoms ($N=0$), so only the basic Gaussian mode has $c_0\neq0$. The density increases to the optimal squeezing density $N^{\mathrm{opt}}_{\mathrm{sngl}}$ at the third step of the plot, and we are able to see the relative contribution of new modes at that coupling strength, observed previously~\cite{MiZhangPRA16}. With further increasing atomic density the contribution of the higher-order modes continues to increase, degrading overall squeezing. 
When the length of the atomic medium is increased at a constant atomic density $N^{\mathrm{opt}}_{\mathrm{sngl}}$, one observes a distinctly different behavior: as the interaction length increases, the strength of the ``contaminating'' modes ($p>0$) goes down, potentially improving the squeezing. Such different effects for increasing the nominal optical depth $NL$ should not be too surprising. The underlying interaction is nonlinear and strongly intensity-dependent, which means that in our case of a tightly focused pump beam the different sections of the beam along its propagation affect the spatial beam profile differently, in a rather complex way. 

% unfolded two pass
\section{Two-cell configuration}
\begin{figure}[h]
	\includegraphics[width=0.5\textwidth]{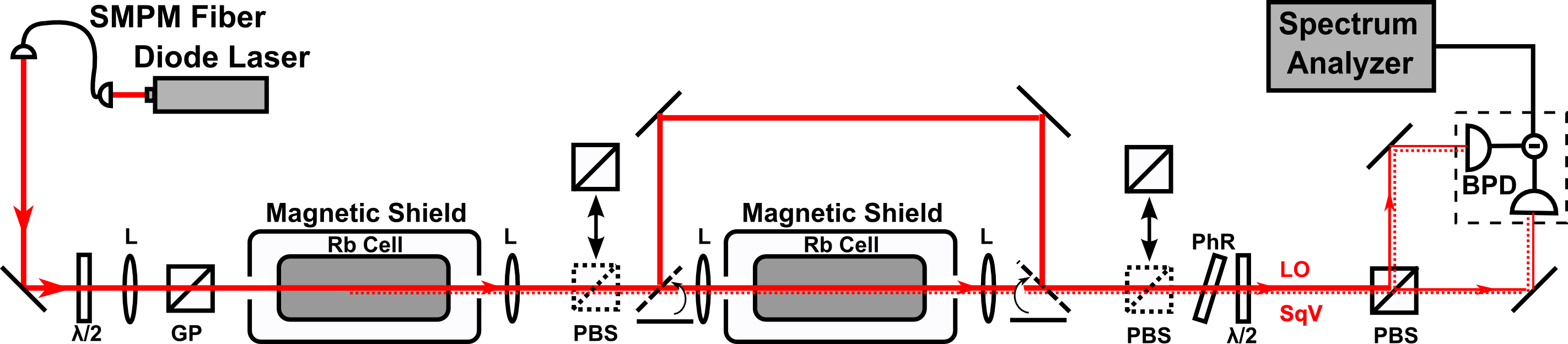}
	\caption{
		\label{fig:setuptwo}(Color online) A modified experimental setup with two independent vapor cells. All the abbreviations are the same as in Fig.~\ref{fig:setup}}.
\end{figure}

To further investigate the effect of multiple interaction of the pump and squeezed vacuum fields with the atoms, we used a modified experimental setup, shown in Fig.~\ref{fig:setuptwo}.  Namely, we added another identical Rb vapor cell in the setup to make a ``unfolded'' double pass.  Such a configuration allows us to avoid some of the limitations of the previous setup, such as, e.g., the inability to independently control the relative position of the beam focus inside each cell. The use of two different vapor cells also allowed us to independently vary their temperatures and relative positions. In addition, it was possible to insert the polarizing beam splitter after the first cell, effectively removing any quantum noise modifications, but preserving any changes in the intensity distribution of the pump field resulting from the interaction with atoms in the first Rb cell. Consequently, for all the experimental parameters we performed two sets of measurements: one with the pump and the vacuum fields propagating through both cells without modifications (which is roughly equivalent to the ``folded'' double-pass measurements), and a ``first-cell-filtered'' case, in which any quantum noise modification in the first cell was rejected and replaced with a coherent vacuum, so the measured squeezing reflected only the pump field modifications in the first cell.

\begin{figure}[h!]
	\includegraphics[width=1.0\columnwidth]{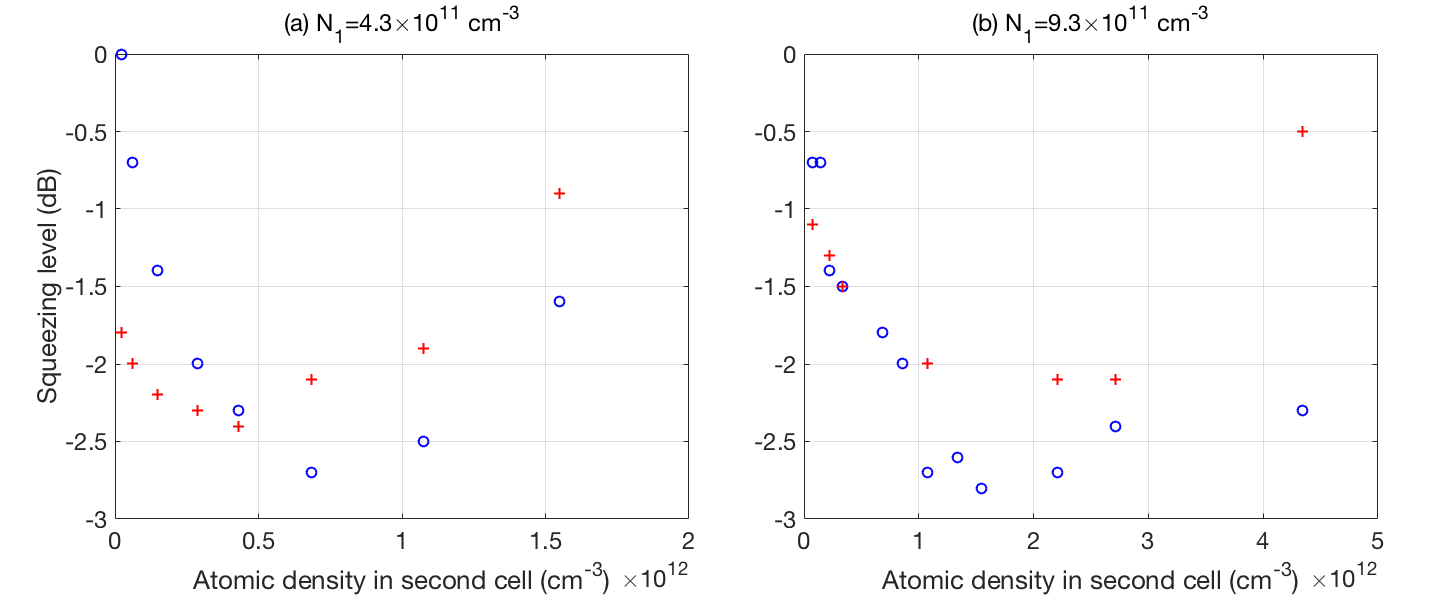}
	\caption{
		\label{fig:density}(Color online)
			A measurement of squeezing dependence on the second cell density in the two-cell system. In (a) the atom density of the first cell is $9.3\times10^{11}$~cm$^{-3}$ and in (b) it is $4.3\times10^{11}$~cm$^{-3}$. Crosses represent the squeezing amount in total and circles correspond to the case when a PBS is inserted after the first cell so that the first-cell-generated squeezing is filtered. 
		}
\end{figure}

Since it was possible to change the temperature of each cells independently, we studied the effect of changing atomic density in the second cell when the the first cell atom density was fixed  at either $N^{\mathrm{opt}}_{\mathrm{sngl}} =9.3\times10^{11}$~cm$^{-3}$, the optimal condition for the single-pass squeezing generation, or at $N^{\mathrm{opt}}_{\mathrm{dbl}} = 4.3\times10^{11}$~cm$^{-3}$, at which the highest value of the measured squeezing in the double-pass configuration was observed. The resulting measurements are presented in Fig.~\ref{fig:density}. It is easy to see that in all cases there exists again an optimal atomic density at which the measured squeezing is optimized. If the squeezed vacuum propagates through both cells (crosses), we unsurprisingly observed the best squeezing when both cells were at the same atomic density, $N^{\mathrm{opt}}_{\mathrm{dbl}} = 4.3\times10^{11}$~cm$^{-3}$, replicating the optimal conditions observed for the double-pass experiment, shown in Fig.~\ref{fig:multi_pass_map}(b). If the first cell is tuned to the optimal single-pass condition, the best squeezing is measured at somewhat higher values of the second cell's atomic density, although there is no further improvements compared to the output of the single cell.

We also observed that if the squeezed vacuum field generated in the first cell was filtered by the polarizing beam splitter, we were able to measure the amount of squeezing after the second cell ($-2.8$~dB) exceeding both single-cell or double-cell best squeezing, for both studied atomic densities of the first cell. This observation implies that the pump modifications due to the interaction with the first cell resulted in more favorable conditions for generation of squeezed vacuum in the second cell. Thus, it may be possible to further improve squeezing by actively optimizing the spatial mode decomposition of the pump field. These experiments are currently under the investigation.  

\begin{figure*}[t]
	\includegraphics[width=1.0\textwidth]{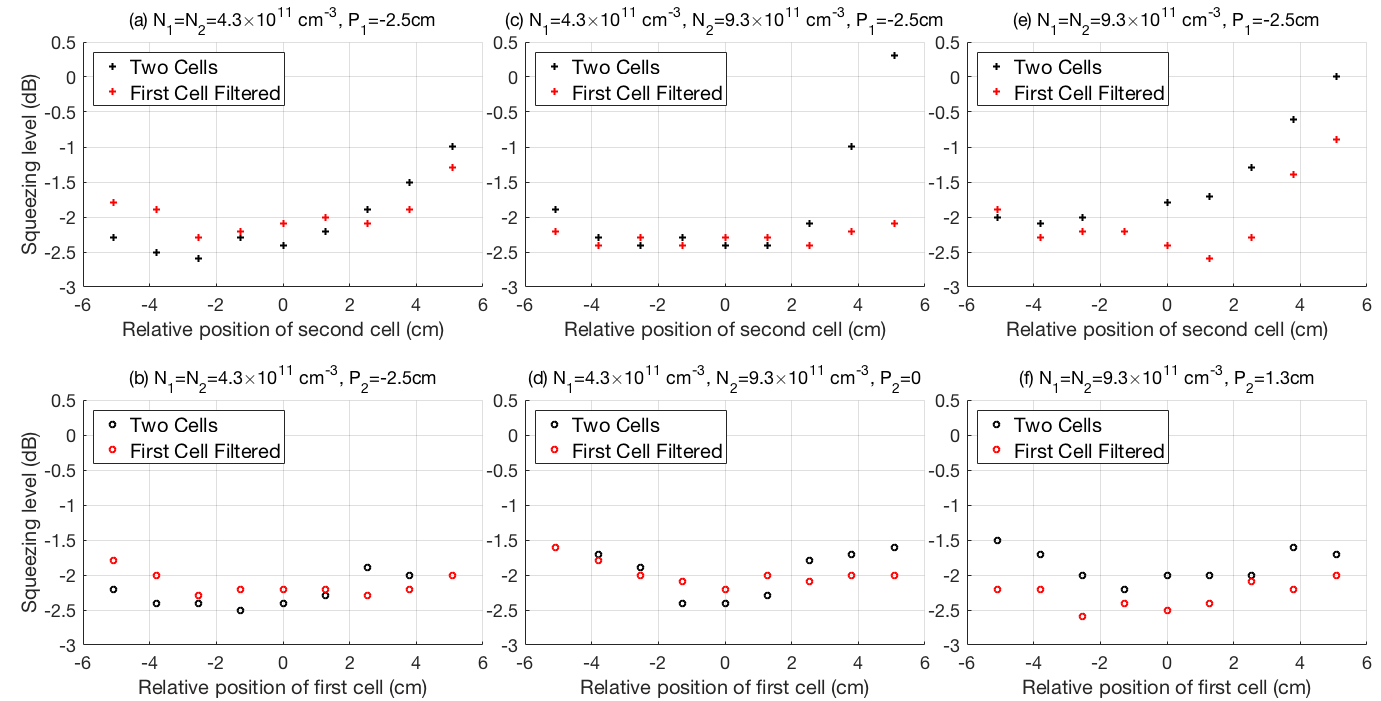}
	\caption{
		\label{fig:position}(Color online)
		The dependence of squeezing amount on position of the first (circles) and second cell (crosses). Black represent the squeezing is measured when both cells are present. Red (light grey) is when an addition PBS is placed after the first cell and filters the squeezed field generated in the first cell.
	}
\end{figure*}

Our experimental results also displayed strong dependence on the observed amount of squeezing to the relative position of the cells, and the focal points of the optical beams. 
Such behavior agrees with our previous observations that the exact decomposition of the higher-order spatial modes is sensitive to the geometry of the interaction, and changes depending on where in the cell the pump beam is focused. 
We changed the relative position of the beam focus and the cell center for each of the two cells. 
We define the relative position of each cell such that for zero the focus of the pump beam is at the geometrical center of the cell, and positive displacement means that the center is after focus, i.e. the beam focus lies in the front part of cell. For example, the displacement of $2.5$~cm means that the focal point is approximately $1.25$~cm away from the front window of our $7.5$~cm cell, while the displacement of $-2.5$~cm places the focal point $5.75$~cm behind the front window, and $1.25$~cm in front of the back window.

For the measurements shown in Fig.~\ref{fig:position}(a), (c) and (e) , we first placed the first (front) cell at its optimal position, where we have best single-cell-squeezing, and optimized the position of the second cell for three different atomic density values.
We see that when both cells had atom density of $N^{opt}_{dbl} = 4.3\times 10^{11}$~cm$^{-3}$ (Fig.~\ref{fig:position}(a)), the best squeezing ($-2.6$~dB below the shot noise) occurred at the second cell position $P_{2} = -2.5~$cm, and corresponded to the squeezed field traveling through both cells. These results are very similar to those obtained in the double-pass experiment. If the squeezing output of the second cell is filtered out, the squeezing level after the second cell drops to $-2.3$~dB.
However, if we increase the atom density in first cell to $N^{\mathrm{opt}}_{\mathrm{sngl}} = 9.3\times 10^{11}$~cm$^{-3}$ while keeping the second cell at $N^{\mathrm{opt}}_{\mathrm{dbl}}$, the best squeezing of $\approx -2.4$~dB occurs at $P_{2} \simeq 0$ (Fig.~\ref{fig:position}(c)); filtering or not filtering the squeezed output of the first cell did not seem to have an effect on the second cell's quantum noise output. Moreover, when both cells were at the atomic density $N^{opt}_{dbl}$, the best squeezing occurs at $P_{2} = 1.3~$cm (Fig.~\ref{fig:position}(e)) for the filtered first cell case, and if the vacuum field traveled unperturbed through both cells, the measured value of squeezing ($-2$~dB at $P_2 \simeq -3.5$~cm) was higher than for the filtered case. 

Similar behavior was observed when we fixed the position of the first cell, and went through similar optimization steps for the second cell position. The measured results are plotted in Fig.~\ref{fig:position} (b), (d) and (f). While we observed less sensitivity to the first cell position and less difference between the filtered and unfilted first cell output, it is clear that to maximize output squeezing the pump beam focus in each consecutive cell must be carefully optimized for each value of atomic density, and deviating from this optimal position can quickly deteriorate the output squeezing. This observation can partly explained why we did not see improved squeezing when we increase the number of passes from two to four in the previous experiment, since no independent focal position optimization was possible for each pass.

\section{Conclusion}
In conclusion, we investigated the behavior of PSR squeezing in case of single or multiple interaction of light with Rb vapor. More specifically, we compared the achievable amount of squeezing after the pump beam traveled through the vapor cell once, or the light after the cell was retroreflected, and the squeezing was measured after the second pass through the cell. As expected from the general principle of nonlinear interactions, the amount of observed squeezing strongly depended on the optical depth of the atomic medium. However, we found, that even though the optimal squeezing was observed at the same value of the optical depth for the single and double pass configurations, the overall value of squeezing was better for the double-pass configuration, when the cell was maintained at the lower temperature. The quadruple pass configuration also yielded the optimal squeezing at approximately same value of the optical depth, although less squeezing was observed compare to the double pass, probably due to lack of proper beam shape control in this complicated geometry. 

We also studied the squeezing in the case of two independent cells, and observed strong sensitivity of the improvement of squeezing in case of the double interaction at lower atomic density. These observations are consistent with our recent finding that the deterioration of squeezed vacuum with optical depth can be (at least partially) explained by the increasing contributions of higher-order spatial modes. Since the process of their generation seems to be less effective in case of extended length of the optical medium compare to the atomic density increase, a possible avenue toward further improvement of squeezing is squeezing generation in a low-finesse cavity, in which the pump field interacts with atoms multiple times, at lower density of atoms. Our current results also add an interesting twist to a long discussion on the equivalence of achieving the high optical depth by increasing the number of atoms or by extending the length of the interaction medium. We showed that in the spatially multimode nonlinear interactions these two methods may lead to different results, and thus one of them can be preferable for a particular application.

In addition, our measurements indicated that it may be possible to further improve squeezing by tailoring the pump spatial profile before the interaction. This conclusion is based on our observation of improved squeezing obtained using the pump field after its interaction with atoms once, as compare to the ideal gaussian spatial profile. We are currently investigating this possibility.

%\section{Acknowledgments}
The authors thank  G. Romanov and T. Horrom for the assistance
with the experiment.  This research was supported by AFOSR grant FA9550-13-1-0098. R. N. L., Z. X. and J. P. D. acknowledge additional support from the ARO and the NSF, and the Northrop Grumman Corporation.

%\bibliography{bibliography/bibliography}

%%%%%%%%%%%%%%%%%%%%%%%

\end{document}